# Bubbles and Market Crashes

Michael Youssefmir, Bernardo A. Huberman, and Tad Hogg

Dynamics of Computation Group
Xerox Palo Alto Research Center
Palo Alto, CA 94304

## Abstract

We present a dynamical theory of asset price bubbles that exhibits the appearance of bubbles and their subsequent crashes. We show that when speculative trends dominate over fundamental beliefs, bubbles form, leading to the growth of asset prices away from their fundamental value. This growth makes the system increasingly susceptible to any exogenous shock, thus eventually precipitating a crash. We also present computer experiments which in their aggregate behavior confirm the predictions of the theory.

## 1 Introduction

Perhaps one of the most fascinating behaviors of markets is the situation in which valuations placed on a particular asset become “self-fulfilling”. One such scenario arises when prices not only reflect the individual market participant’s valuation of the asset (the fundamentals), but also reflect an extra amount that is often called a “bubble”. This extra valuation may arise because market participants believe that they will be able to sell the asset in the future to other participants at a higher price above and beyond what they actually think it is worth. Although, there is still controversy as to the existence of such bubbles, some of the more famous examples that are cited are the Dutch Tulip Bulb mania, the South Sea Bubble in England, the rise and subsequent collapse of the Mississippi Company in France, and the 1929 and 1987 stock market booms and crashes. Typically, during an initial period, large price increases can be justified by fundamental valuations placed on the asset. However, these price movements begin to be extrapolated by the market participants and further price movements occur in anticipation of this. In this way, the price movements become a self reinforcing process independent of any valuation placed on fundamentals [1], and a bubble forms. As the process continues, market participants with an eye to the fundamentals start to withdraw, expected price movements no longer occur, and the premium placed on these expected price movements rapidly disappears resulting in a crash or bursting of the bubble.

If one accepts that such phenomena are not related to the market fundamentals then it is difficult to gain an understanding of “bubbles” within the standard rational expectations theory. It has been shown, for example, that if rational agents have long horizons into the future, such bubbles simply cannot occur (see for example [2, 3]). The argument is that rational agents in this case know that prices are not warranted by fundamentals and can wait until the bubble bursts. These agents, expecting the bubble to eventually burst, will move against the bubble before this event. But this movement itself will cause the bubble to burst; and, thus, tracing back through the argument, it is clear that under standard rational expectations theory such bubbles cannot exist [4].

Part of the reason for the controversy over the existence of bubbles is that the fundamentals are themselves uncertain. The fundamental value of an asset can be obtained by discounting the future earnings stream of the asset along with the future terminal value. Indeed, it is the intrinsic uncertainty in estimating the earnings stream and proper future discounting that makes the detection of bubbles so hard and their existence controversial. Arguments have been made, for example, in which the Tulip mania, the South Sea Bubble, and the Mississippi bubble can be explained by rationally justifiable expectations from the point of view of the market participants at the time [5]. Moreover, various models have shown that price paths that can be interpreted as a bubble are qualitatively equivalent to those from a rational expectations viewpoint, in which certain expected events did not occur (so called “omitted variables”) [6–8].

However, it is not as easy to provide reasonable explanations of the 1987 Stock Market Crash that can account for the devaluation of U.S. corporate stocks by about 22% on October 19, with no real discernible news [9].

Instead of focusing on rational expectations models, others [10–12] have focused instead on a “noisy” trader approach to modelling financial markets that stress psychological forces in investor behavior[13]. In this approach, traders are not fully rational and their actions and beliefs are subject to systematic biases that are decoupled from the actual fundamentals. These traders are termed as “noisy”. Psychological experiments as well as various market survey results are cited to suggest that such noise traders have trend chasing tendencies that are decoupled from fundamentals. For example, in one such survey [14], it was found that professional currency forecasters during the mid-1980s expected price trends to persist over a short horizon, but, at they same time, they expected a long run return to fundamentals.

It is also argued that the rational traders themselves may be limited in their ability to counteract such systematic biases, or, in other words arbitrage against the mispricings of the fundamentals. One such limitation could be the result of uncertainty about the fundamentals, and it is therefore never fully certain that an arbitrage opportunity even exists. There is also uncertainty about the time horizon over which such systematic biases can last, and unless traders have very long horizons, they again cannot fully take advantage of the arbitrage situations. In this way, the arbitrage opportunities will carry some risk, and risk-averse arbitraguers may not be able to fully arbitrage against “noisy” traders. It may in fact be quite profitable not to trade against a wave of self-reinforcing price movements, but, rather, to jump on the bandwagon, jumping off only near the top. Such a trading strategy is exemplified by view of market mechanisms presented by George Soros, who stresses the importance of investor bias’s and self reinforcing trends in security analysis [15].

Central to the theme of the present analysis is the idea of “positive feedback” traders as discussed in [12]. Positive feedback traders, are those who are more inclined to buy as prices rise and sell when prices fall. Otherwise known as trend chasing, positive feedback trading takes on many forms commonly found in financial markets, such as stop-loss orders, margin call liquidation, and simple extrapolative expectations. These trading strategies provide a reinforcing mechanism for market trends which can lead to large swings in the value of the asset traded. Reference [12] presents some of the indications that positive feedback strategies are indeed common in financial markets, as well as some survey and experimental evidence to that effect.

For example, psychological experiments [16] have been conducted in which subjects are shown real stock prices from the past and asked to forecast subsequent changes, while performing trades consistent with these forecasts, and, by so doing accumulate wealth.

These subjects, of course, were asked to trade only based on past prices and were not exposed to external "fundamental" news. It was found that subjects track the past average when the stock prices are stable, thus trading against price fluctuations when they arise. However, as prices began to show consistent trends, they began to switch to a trend chasing strategy, buying more when prices increase and selling when prices decrease. Perhaps, even more compelling evidence of the presence of trend chasing strategies, is the wide prevalence of "technical analysis" that tries to spot trends and trend reversals by using technical indicators associated with past price movements [17, 18].

In contrast to the trend chasing behavior of positive feedback traders, fundamentalist trading refers to buying and selling decisions that are made with a view to the underlying value of the asset. In the equity markets, for example, this fundamental value is obtained by discounting the future dividend stream of the asset. Fundamentalist traders tend to push prices back towards fundamentals by trading equivalent securities against each other or by simply selling when prices are high and buying when prices are low. We use "fundamentalist" synonymously with "arbitraguer", since both trading behaviors involve a valuation placed with respect to the underlying value of an asset. In this respect, their buying and selling decisions are influenced by past price movements only in the way such price movements affect the underlying valuation of the asset; they do not extrapolate trends.

In this paper, motivated by the idea of "positive feedback" trading and trend chasing biases, we investigate the types of behaviors that may be expected when market participants do indeed have the type of extrapolative expectations described above. More specifically, we assume that individuals tend to extrapolate trends for a period of time we call the trend horizon, but nevertheless expect prices to return to some fundamental value in the long run. This type of expectations is, for example, consistent with the survey results of [14], in which currency forecasters during the mid 1980's were asked about their expectations about future price movements. In the model presented in this paper, when trend horizons are small, market prices tend to relax to a stable value at the average of the fundamental valuations placed on the asset by market participants. However, as the individuals' trend horizons get larger, this fixed point eventually becomes unstable.

It might be assumed and it has been argued that the destabilizing effect of "positive feedback" traders may disappear from markets as such traders lose their wealth to the rational fundamentalist traders [19]. However, as [20, 21] point out, this may not be entirely clear. Positive feedback traders tend to take on more risk than they might think they have taken on, and if this risk is rewarded, their performance will turn out to be better than they might have otherwise thought. Furthermore, from the point of view of our model of bubbles and crashes, bubbles tend to be one time phenomena that occur with infrequently ; so that, in this way, such trend chasing agents will learn about the error of their strategies only once the bubble has come and gone. Indeed, if trend extrapolation is initially validated by rising fundamentals, agents might incorrectly form longer trend horizons that will sustain rising prices and a bubble even if fundamentals no longer justify these increases.

The approach taken in this paper allows for an interplay between fundamentalist and positive feedback trading and the relevance to the formation and breaking of bubbles. Indeed, we do not even make a sharp distinction between these two forms of expectations, but we construct a model of expectations that smoothly interpolates between the two types of behaviors. The central result that comes out of our simplified model concerns the degree to which bubbles themselves are susceptible to external shocks. Through a theoretical analysis of the dynamics, we find that bubbles become more and more susceptible to external shocks as they develop, until either a shock manages to burst the bubble or the bubble itself runs its course and disappears.

## 2 The Model

In this section, we present a model consisting of heterogeneous agents participating in an asset market. We stress that we are making an effort to combine the simplest model of market dynamics with the type of expectations that have been observed and surveyed in the financial markets and psychological experiments. In order to make it analytically tractable and since we are dealing with a dynamic theory of price movements, we have to make several assumptions and simplifications about the market mechanisms involved. We assume that only a few individuals commit to buying or selling at any given time. Agents become market participants at some rate given by a Poisson process with rate $\alpha_w$, and thus have an effect on the prices of the given asset. In this paper, we also make a simplifying assumption that this rate is constant in time and that all agents enter similar trades. This means that within the context of the model, that the volume of trades is on average a constant and that price movements caused by single agents tend to be equal in magnitude.

Once an agent participates in the market, a decision is made to be a buyer or seller, based on beliefs and expectations on the future direction of the price. When an agent enters the market, the price of the asset either increases or decreases by an amount proportional to a quantity, denoted by$\alpha_p$, and inversely proportional to the total number of agents, $N$, in the market. This assumption ignores the dynamics of the underlying auction or market clearing mechanism. However, it does encapsulate in the simplest way possible the idea that prices tend to rise (or fall) when there are more (or less) buyers than sellers.

We assume that individuals' fundamental beliefs about the value of the particular asset is intrinsically "noisy". That is individuals not only disagree about the fundamental value of the asset, but are also affected by the beliefs of others as they are reflected in price movements. More concretely, each agent has determined a fundamental value, that the given asset is worth. We call this fundamental price, $p_f^i$, for agent $i$. We further assume that these fundamental asset prices are distributed as a Gaussian random variable with average value, $p_f^{av}$ and variance, $\delta$.

### Expectations

The decision to buy or sell is based upon an individual's expectation of the future price of the asset. We assume that these expected price movements are discounted over a horizon, $H_w = \frac{1}{\alpha_w}$, representing the average amount of time that a particular agent spends between decisions to buy or sell. In this way, agents that are more concerned about near term price movements wake up more rapidly, and agents that are more concerned about long term price movements of the asset wake up less often. For simplicity, we assume here that all agents discount over the same horizon.

By their very nature, fundamentalists expect prices to return to their fundamental value in the very near future, while speculators will tend to follow the prevailing trend. Speculative expectations are formed by simply observing the trend in the market price of the asset over a backward horizon, $H_b$. Denoting this trend over the past at time $t$, by $T(t)$, then

$$T(t) = \frac{1}{H_b} \int_{-\infty}^{t} \frac{dp}{d\tau} \exp\left(-\frac{(t-\tau)}{H_b}\right) d\tau. \quad (1)$$

This is an exponentially smoothed average rate of change of prices over a past horizon $H_b$[1]. We make the simplifying assumption that this past horizon is the same for all agents.

In order to describe the two sometimes conflicting tendencies between speculation and fundamentalist trading, we parametrize an agent's particular tendency by a trend horizon $H_t$, over which trends are expected to last. Thus, an agent inclined towards fundamentalist trading will be characterized by a very small $H_t \ll H_w$, whereas a speculator will be one for which $H_t \gg H_w$. Intermediate values of $H_t$ denote agents with characteristics intermediate between fundamentalists and trend chasers. Notice that in the fundamentalist case there will be a tendency for prices to move towards their fundamental value with trends playing a minor role, whereas speculators will base a buy/sell decision primarily on the current trend.

[1] An integration by parts will show that $T(t) = \frac{p(t) - \langle p(t) \rangle_{H_b}}{H_b}$, where the angled brackets denote an exponential average over the horizon $H_b$, so that $T(t)$ measures the deviation from the moving average of prices. This type of measure is indeed one of the most popular among technical analysts for measuring underlying trends.

Given these trading characteristics, the rate at which prices will be expected to change will be made up of three components. Denoting, the expected price at time, $t+\tau$, by $p'(t+\tau)$, we can write an equation for average expectations about price movements:

$$\frac{dp'}{d\tau}(t+\tau) = \frac{-(p'-p_f)}{H_t} + \left(T(t) + \frac{(p(t)-p_f)}{H_t}\right)\exp\left(-\frac{\tau}{H_t}\right). \qquad (2)$$

Together with the boundary condition $p'(t+\tau)_{\tau=0} = p(t)$, this equation gives the average price expectations of agents parametrized by fundamental valuation placed on the asset by the agent, $p_f$ and trend horizon, $H_t$. Note again that $p_f$ here is randomly picked for each agent from a Gaussian distribution with average value, $p_f^{av}$, and variance, $\delta$.

The terms in this equation are justified as follows: first, the first term on the right hand side reflects the expectation that prices will eventually relax to their fundamental values at a rate given by

$$\frac{dp'}{d\tau} \underset{\tau\to\infty}{\to} -\frac{(p'-p_f)}{H_t}. \qquad (3)$$

Second, the observed trend, $T(t)$, is expected to last for the near term, but have a decreasing influence over a time span of the order of the trend horizon, $H_t$. The second term in (2) does not corrupt relation (3) but insures that

$$\frac{dp'}{d\tau} \underset{\tau\to 0}{\to} T(t), \qquad (4)$$

so that the trend is indeed expected to continue in the very near term. The push and pull between these two terms will mean that agents expect a persistence of the trend over the trend horizon, and also a gradual return to fundamentals over a period of the order of a few trend horizons.

Finally, we add to the right hand side of eqn. (2) a term that is variable from agent to agent and that reflects the diversity and variability in beliefs about future price movements. We model this intrinsic uncertainty in individual expectations by a random modulation of the future trend through the delta correlated Gaussian process, $\xi(\tau)$ that satisfies

$$\langle\xi(\tau)\rangle = 0 \quad and \quad \langle\xi(\tau)\xi(\tau')\rangle = \sigma^2\delta(\tau-\tau'). \qquad (5)$$

This term introduces stochasticity into the individuals' expectations.

We are now in a position to write an equation that governs the way particular agents expect prices to move. It is given by the following stochastic equation

$$\frac{dp'}{d\tau}(t+\tau) = \frac{-(p'-p_f)}{H_t} + \left(T(t) + \frac{(p(t)-p_f)}{H_t}\right)\exp\left(-\frac{\tau}{H_t}\right) + \xi(\tau). \qquad (6)$$

together with the initial condition $p'(t+\tau)_{\tau=0} = p(t)$. The solutions of this equation, with given initial condition, determine the distributions of price expectations for individuals whose speculative characteristics are determined by the value of the trend horizon, $H_t$.

A few remarks are in order. First, this model produces an ensemble of price movements, and particular agents can be thought of as picking the future expected price movements from this ensemble. Second, in the absence of speculation, the trend horizon becomes very small and the dynamics is determined by the first term of Eqn. 6, which exhibits exponentially fast relaxation to the fundamental price.

In order to illustrate perceived expected prices generated by our model, we solve eqn. 6, and plot in fig. 1 the average behavior of these expected prices for different current prices and trends, when the fundamental valuation is at 2000. The cases shown use the above expectations with $H_t = 200$. The three cases shown have the three current prices set at 2000, 1900, and 1800, with trends, $T$ given by +1,-1, and 0 for each case. Note that for the current price set at the fundamental value of 2,000, the reversal in price and direction of price change occurs in about 200 time steps, the value of $H_t$. Furthermore as prices draw away from the fundamental, this reversal tends to occur earlier even though $H_t$ itself is kept constant.

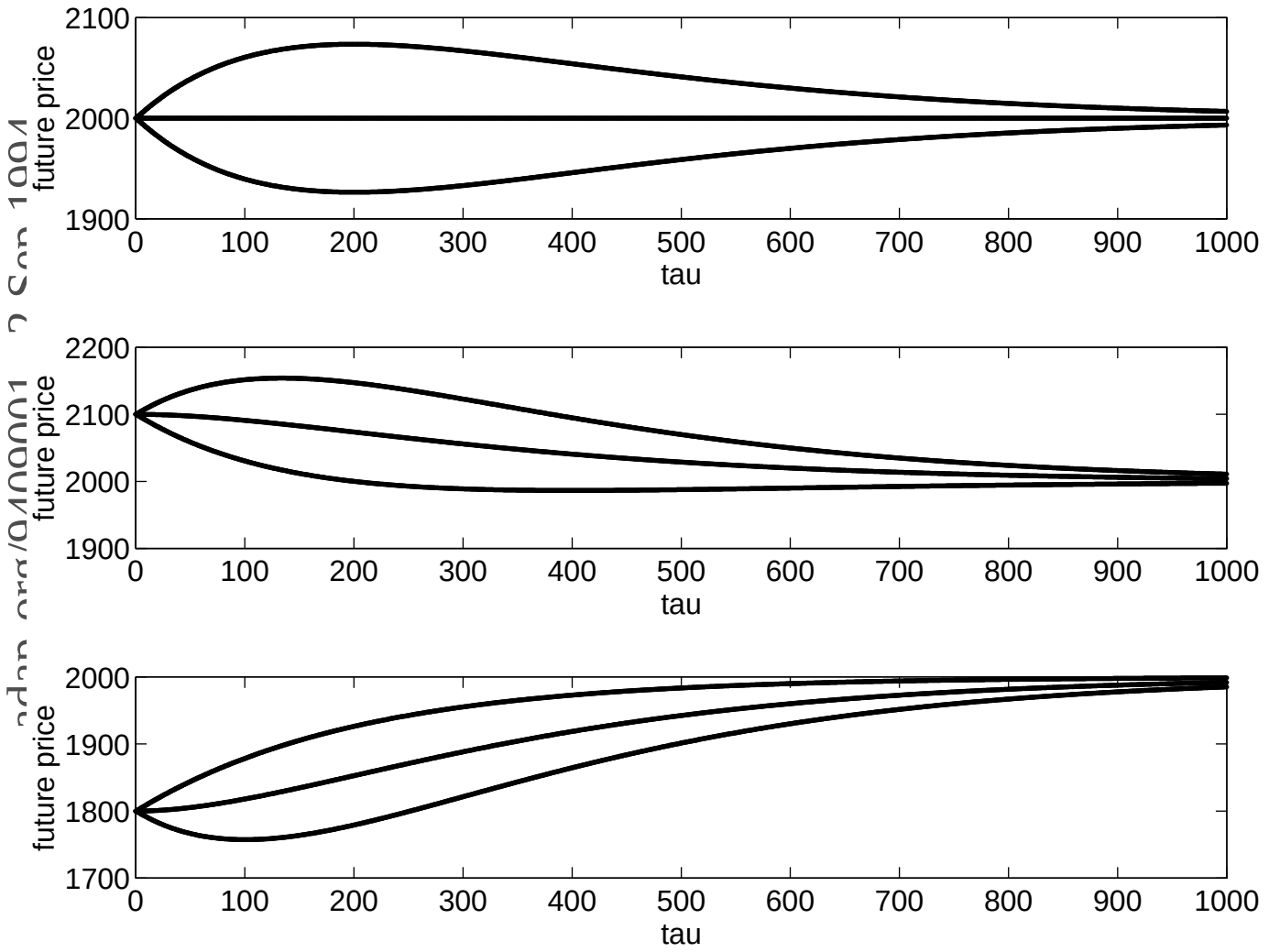

**Fig. 1.** Expectations of individuals about future price movements given by integrating eqn. 6 for the case in which $\sigma = 0$ and $H_t = 200$. Fundamental valuation $p_f$ is set at 2000. The three cases show the expectations when the current market price, $p'(t+\tau)_{\tau=0} = p(t)$ is set at 2000 (upper figure), 2100 (middle figure), and 1800 (lower figure). For each figure, the three curves represent the trends, $T(t)$ given by +1, 0, and -1.

## Decision Making

We are now in a position to calculate the probability that individuals make specific trading decisions. Solving eqn. 6 for the future behavior of price movements, we obtain,

$$p'(t+\tau) = p_f + (p(t) - p_f)\exp\left(-\frac{\tau}{H_t}\right) + \left(T(t) + \frac{(p(t)-p_f)}{H_t}\right)\tau\exp\left(-\frac{\tau}{H_t}\right) + \int_0^{\tau} \xi(\tau)d\tau. \tag{7}$$

If we now discount this over the wake up horizon, $H_w$, we obtain,

$$\hat{p}(t) = \frac{1}{H_w}\int_0^{\infty} \exp\left(-\frac{\tau}{H_w}\right) p'(t+\tau)d\tau, \tag{8}$$

which represents the value placed by an individual on the asset. If this value is above the current market price, $p(t)$, then the individual is a buyer of the particular asset, and is a seller of that asset otherwise i.e.

$$\text{if} \begin{cases} \hat{p}(t) > p(t) & agent\ is\ buyer \\ \hat{p}(t) < p(t) & agent\ is\ seller \end{cases} \tag{9}$$

Solving for $\hat{p}(t)$, in the above equation shows that it is distributed as a Gaussian with mean,

$$\langle \hat{p}(t) \rangle = T(t)\frac{H_e^2}{H_w} + (p(t) - p_f)\frac{H_e}{H_w}\left(1 + \frac{H_e}{H_t}\right) + p_f \tag{10}$$

and variance,

$$\sigma^2_{\hat{p}(t)} = \sigma^2 \frac{H_w}{2}, \tag{11}$$

where $\frac{1}{H_e} = \frac{1}{H_w} + \frac{1}{H_t}$.

The probability that a particular individual is a buyer can then be written as,

$$Prob(\hat{p}(t) > p(t)) = P(p, T, p_f, H_f) = Q\left(-\frac{T(t)\frac{H_e^2}{H_w} + (p(t) - p_f)\left[\frac{H_e}{H_w}\left(1 + \frac{H_e}{H_t}\right) - 1\right]}{\sigma_{\hat{p}(t)}}\right), \tag{12}$$

where $Q(x) = \frac{1}{2}\left(1 - erf\left(\frac{x}{\sqrt{2}}\right)\right)$. This probability is parametrized by a continuum of possible values, from fundamentalists with very small trend horizons to purely trend followers that have large $H_t$.

Finally, we assume that as prices get further away from fundamentals that agents become more and more inclined towards lower trend horizons. For example, as prices draw away from the fundamental valuation, the risk of arbitrage opportunities become more and more diminished. Agents will be more and more inclined to make their decisions as pure fundamentalists ignoring any prevailing trends. Therefore, in general for each agent, we take $H_t$ to be a monotonically decreasing function of $(p - p_f)$ with an assumed form,

$$H_t(p - p_f) = H_t^0 exp\left(-\frac{(p - p_f)^2}{2\beta^2}\right), \tag{13}$$

so that as prices move away from fundamentals, individuals are less and less inclined to speculate. Here $\beta$ parametrizes the price range over which agents are more prone to longer trend horizons.

| *Symbol* | *Meaning* |
|---|---|
| $\alpha_w$ | Wake up rate |
| $\pm\alpha_p$ | Change in price due to buy/sell trade |
| $p_f^{av}$ | Average fundamental value placed on asset |
| $\delta$ | Variance in fundamental value placed on asset |
| $p_f$ | Fundamental value placed on asset by an agent |
| $H_t$ | Trend (forward) horizon |
| $H_b$ | Past (back) horizon |
| $\sigma$ | Variance of future random process, $\xi(t)$, in expectations |
| $\beta$ | Price range over which agents reduce trend horizons |

**Table 1.** Table showing the parameters in the model.

## 3 Bubble Dynamics

To make headway into the dynamics generated by the above model of expectations, we now write the equations that describe the macroscopic dynamics of the market. The equation governing the evolution of trends is easily obtained by taking the time derivative of Eqn. 1. It yields

$$\frac{dT}{dt} = -\frac{T}{H_b} + \frac{1}{H_b}\frac{dp}{dt}. \tag{14}$$

where, again, we assumed that all agents determine trends over the same back horizon, $H_b$.

In order to obtain an equation for the dynamics of prices in the limit of large numbers of agents, we note that the price increment in a given interval of time $\Delta t$ is given by

$$\Delta p = N\alpha_w \Delta t \left[\frac{\alpha_p}{N}\rho(p,T) - \frac{\alpha_p}{N}(1-\rho(p,T))\right] \tag{15}$$

where $\rho(p,T)$ is the probability that, given a price $p$ and a trend $T$, an agent will become a buyer. This probability is given by

$$\rho(p,T) = \langle P(p,T,p_f)\rangle_{p_f}, \tag{16}$$

where the angled brackets denotes an average over the distribution of fundamental values placed on the asset by the agents. In the continuum time limit ($\Delta t \to 0$), we thus obtain

$$\frac{dp}{dt} = \alpha_p\alpha_w(2\rho(p,T)-1) \tag{17}$$

which, together with Eqn. 14, will be formally simplified to read as

$$\frac{dp}{dt} = \mathrm{P}(p,T) \tag{18}$$

$$\frac{dT}{dt} = -\frac{1}{H_b}(T - \mathrm{P}(p,T))$$

where the function $\mathrm{P}(p,T)$ is given by

$$\mathrm{P}(p,T) = \alpha_p\alpha_w[2\rho(p,T)-1]. \tag{19}$$

These equations describe the macroscopic dynamics of prices and trends, while ignoring the intricacies of individual utilities and beliefs. In order to gain some understanding about the nature of the solutions to these equations, we first look for an equilibrium. The most natural equilibrium is one in which the price is stable at the fundamental value averaged over all the agents in the market, $p = p_f^{av}$. Since the price is stable, there can be no trends, and $T = 0$. Moreover,

$$\mathrm{P}\big(p = p_f^{av}, T = 0\big) = 0 \tag{20}$$

determines the only fixed point, since the expression in eqn. (12) is an odd function around $p = p_f^{av}$ when $T = 0$.

The stability of this fixed point is governed by the eigenvalues of the stability matrix given by,

$$S = \begin{bmatrix} \frac{\partial \mathrm{P}}{\partial p} & \frac{\partial \mathrm{P}}{\partial T} \\ \frac{1}{H_b}\frac{\partial \mathrm{P}}{\partial p} & -\frac{1}{H_b}\Big(1 - \frac{\partial \mathrm{P}}{\partial T}\Big) \end{bmatrix} \tag{21}$$

where all derivatives are evaluated at the fixed point $T = 0$ and $p = p_f^{av}$. If the matrix, $S$, has eigenvalues with positive real part then the fixed point is unstable, while negative real components imply that the fixed point is stable. Figures 2 and 4, show some of the different dynamics that arise out of the above equations.

## Dynamics Without Bubbles

We first present numerical integrations of eqns. (18) for parameter values that exhibit relaxation to the average fundamental valuation placed on the asset by the agents. Here the average wake up time for each agent is set to $H_w = 100$. Figure 2 shows the results of the numerical integration for these parameter values. We have set the value of $H_t = 5$ to be small on the time scale over which agents wake-up, corresponding to the regime in which agents are fundamentalists and not trend chasers. In this rather trivial instance, the price settles at the average fundamental value placed on the asset by the agents, $p_f^{av}$; there is no bubble. We obtain for the stability matrix,

$$S = \begin{bmatrix} -1.0000 & -0.2526 \\ -0.0050 & -0.9676 \end{bmatrix} \tag{22}$$

which has eigenvalues —0.1120 and —0.0048.

In order to gain more intuition about these dynamics, we can treat the pair $p$ and $T$ as a vector and draw the flow field that describes its dynamics. The flow vectors on this plot, describe the direction and magnitude of the evolution of the system given a price and a past trend. This is shown in figure 3. As expected, flow vectors are pointed in such a way that prices relax towards $p = p_f^{av}$ and $T = 0$.

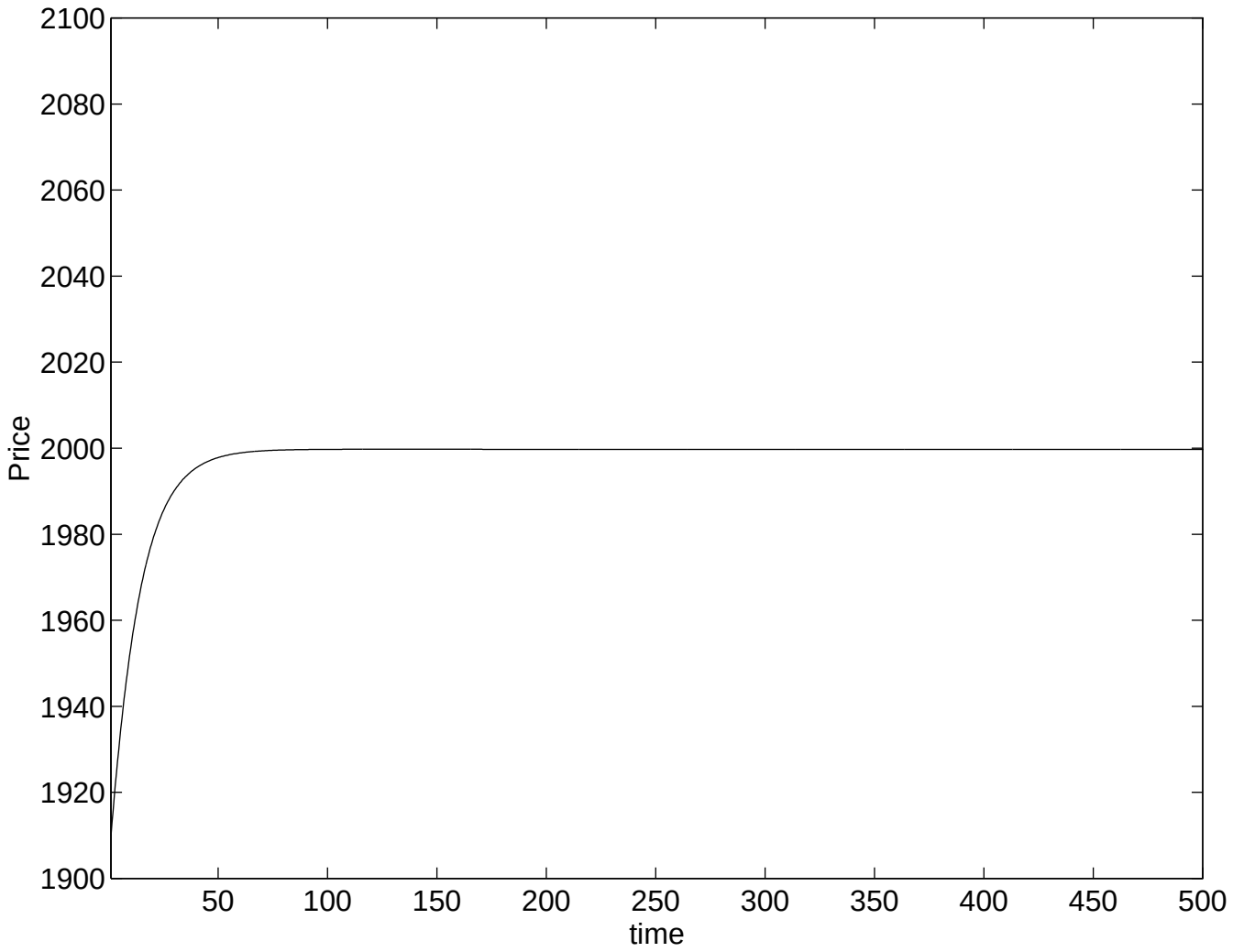

**Fig. 2.** The dynamics generated by eqns. 18. The parameters here are , $\delta = 100$, $\beta = 200$, $p_f^{av} = 2000$, $\alpha_w = 0.01$,$\alpha_p = 1000$, $\sigma = 1$, $H_b = 210$, and finally $H_t = 5$. Note here that agents here tend to extrapolate trends only over short time periods compared to their wake up times. Hence the dynamics quickly relaxes to the average fundamental valuation placed by the agents.

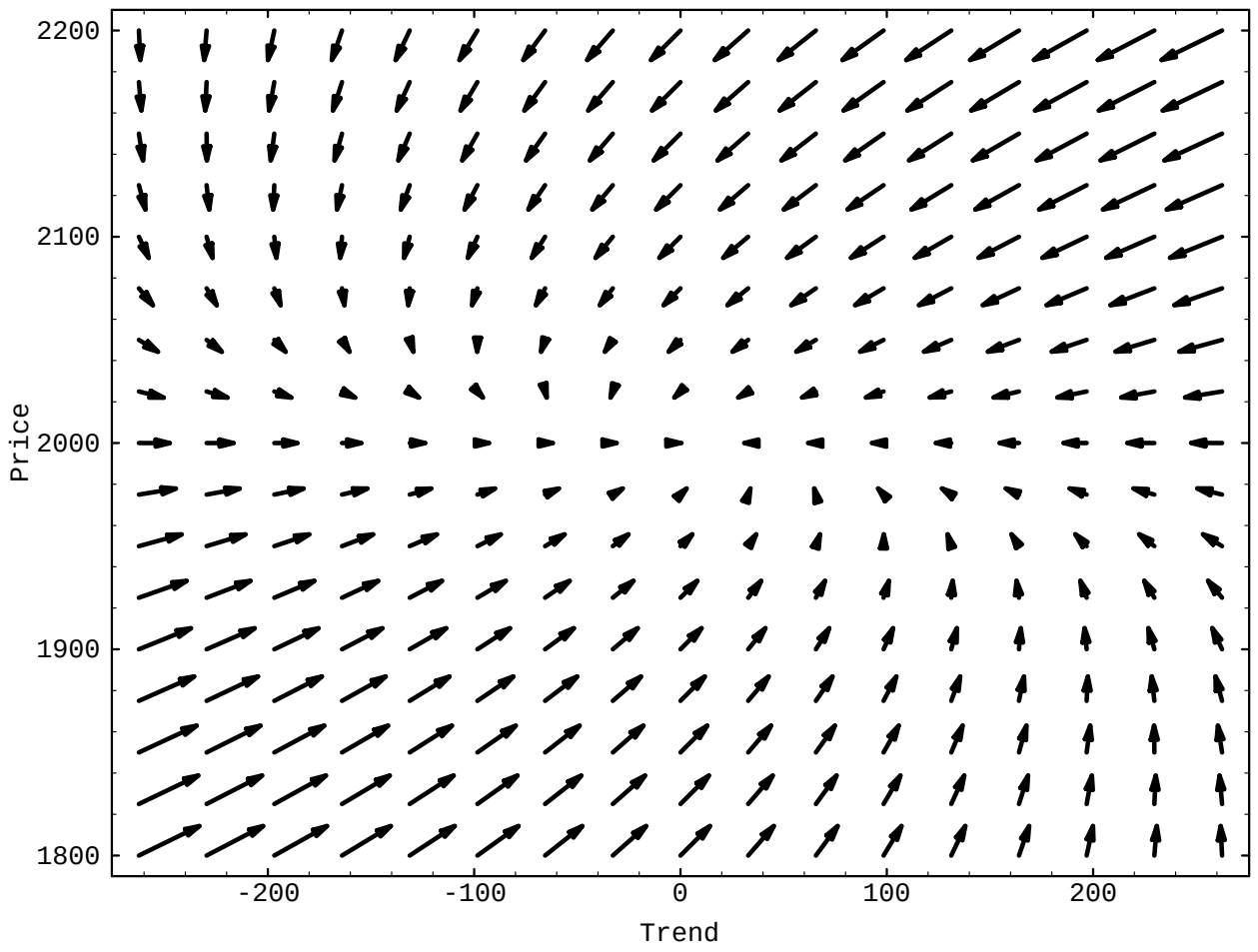

**Fig. 3.** Flow field for the price, $p(t)$, and past trend, $T(t) * H_b$, where the multiplication by $H_b$ is performed so that both axes have equivalent units. Parameters are the same as in fig. 2 and the price and trend are quickly pushed towards equilibrium values at the center.

## Bubbles

In the second example, figure 4, individuals now have trend horizons that are long enough so that the fixed point, $p = p_f^{av}$ and $T = 0$, becomes unstable. Prices now rise beyond $p_f^{av}$, level off, and then begin a slow and then sharp decline to prices below $p_f^{av}$. After this, the cycle then begins again and rises up from below $p_f^{av}$, through this value, and so on. In this regime the extrapolative expectations support price levels beyond the fundamental, simply because trend chasing by individuals accentuates price movements, after which other trend chasers are self consistently more likely to act based on this trend. Again the stability of the fixed point can be obtained from the eigenvalues of the stability matrix Eqn. 21. The eigenvalues are both positive (0.0064 and 0.0578) and the fixed point is therefore unstable.

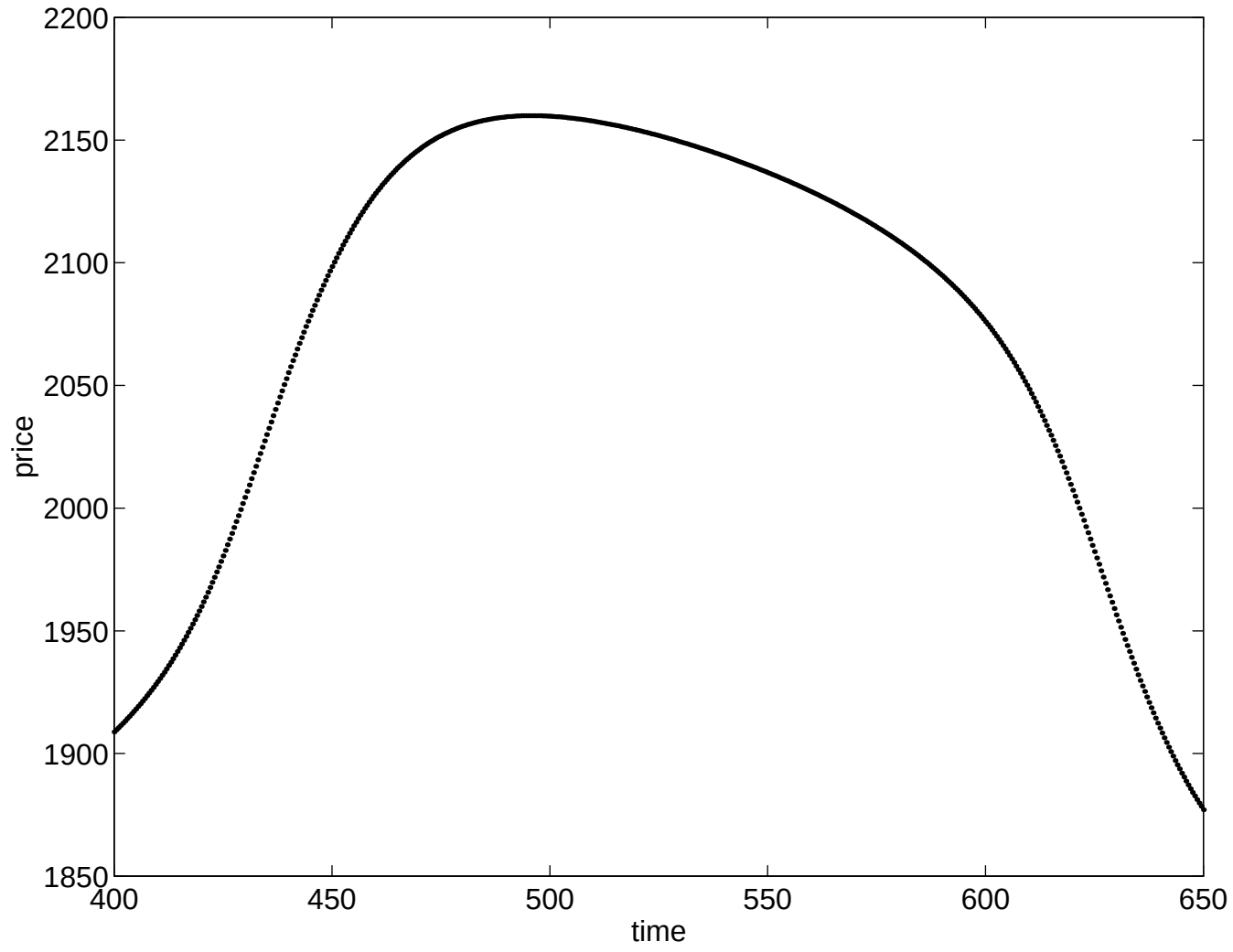

**Fig. 4.** The dynamics generated by eqns. 18 for same parameters as fig. 2 except that $H_t = 200$. Note here that agents here now extrapolate trends far enough in their expectations that the fixed point in fig. 2 is no longer stable. A bubble is formed and eventually bursts.

Figure 5 shows the corresponding vector field. The flow field is now such that the system executes periodic motion around the fixed point. As can be seen from the flow field, when the price is at $p_f^{av}$, and the trend is positive, the dynamics flow in the direction of increasing price. The point here is that prices will not immediately head back toward $p_f^{av}$, but must first rise, reach a critical point, and then flow back to the lower levels. In this way the bubble is naturally self sustaining.

Here it should be pointed out that our model of agent expectations is limited in that agents in the model do not learn. Indeed, historical examples of bubbles do not exhibit periodic formation and crashing of the bubble as the above equations would imply. Agents that buy at the top of the bubble in this model, do not learn from their mistake and do so again and again, leading to the cyclic nature of the solutions in the above equations. However, one would indeed expect that, after the bubble bursts, that actual market participants will recognize the error of their ways and are then less prone to trend chasing. The model as such, however does capture the essential dynamics of the formation and consequent collapse of the price bubble.

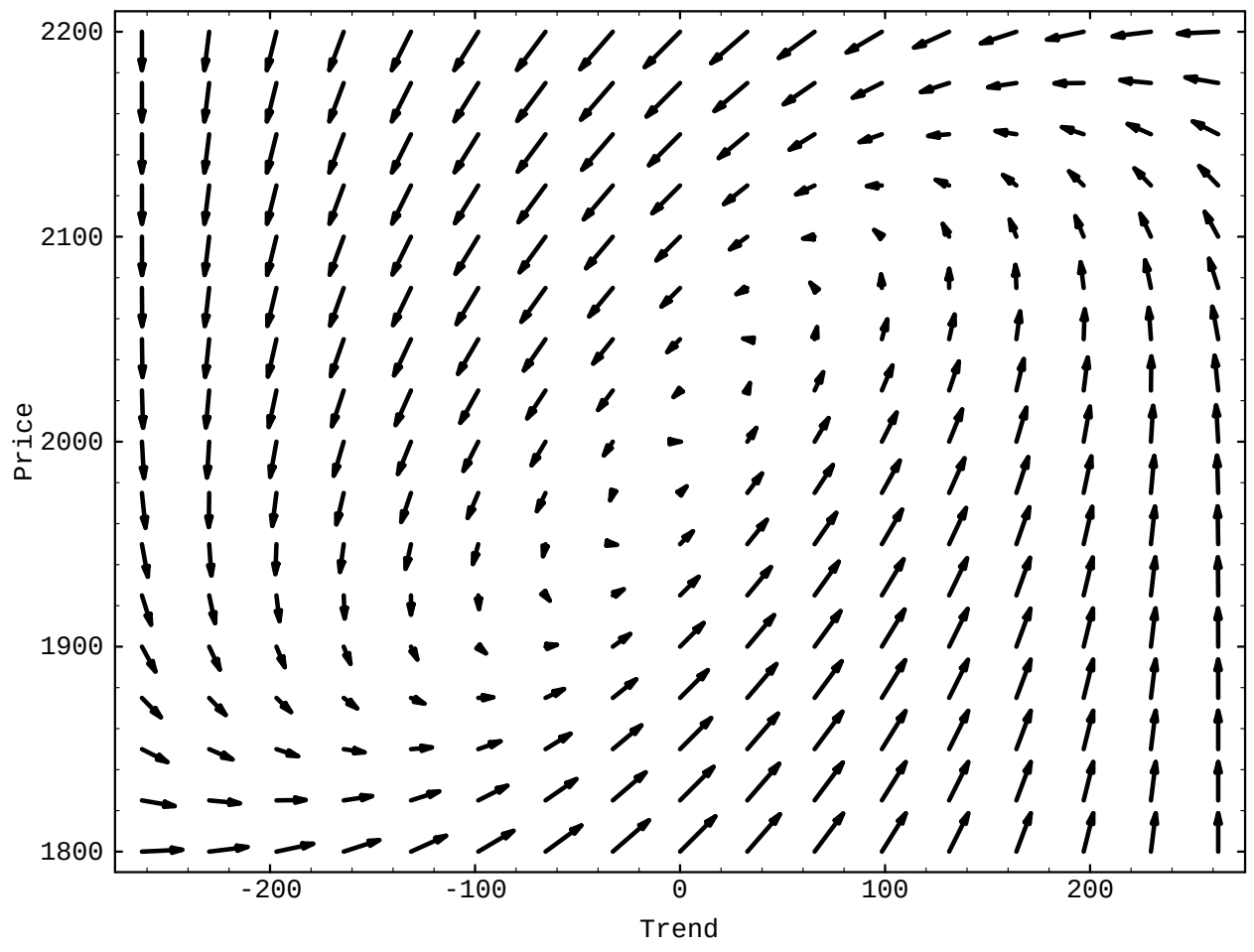

**Fig. 5.** Flow field for the price, $p(t)$, and past trend, $T(t)$. Parameters are the same as in fig. 4 and as can be seen bubble formation takes place as the system moves up the right side of the flow field. It reaches a critical point after which the system flows back down the left side of the flow field and the bubble disappears.

# 4 Computer Experiments

Eqn. 18 is only an approximation to the dynamics of prices and trends in the sense that it becomes accurate in the limit of very many agents. In order to show the validity of this approximation, and in order to investigate effects associated with only a finite number of agents, we ran simulations of the model proposed above. To this end, and for clarity, we state the exact nature of these computer simulations:

In these experiments each agent is randomly assigned a "fundamental valuation", $p_f^i$, where $i = 1, ..., N$ labels the agents. These fundamental valuations are drawn from a Guassian distribution with average $p_f^{av}$ and variance $\delta$.

Next each agent randomly wakes up at a Poisson rate, $\alpha_w$, at which time a buy or sell decision is made. Under the assumptions of our model, an agent is a buyer at time $t$ if the random variable $\hat{p}(t)$ is greater than the current price, $p(t)$, and is otherwise a seller. $\hat{p}(t)$ is distributed as a Guassian random variable with mean given by eqn. 10, and variance by eqn. 11.

Finally, when an agent wakes up and becomes a buyer, the market price increases by an amount $\frac{\alpha_p}{N}$, and it decreases by $\frac{\alpha_p}{N}$ if the agent becomes a seller.

The experiments typically comprised a large number (10,000 or larger) of agents, with given horizons and fundamental values entering as parameters. For these parameters, the unit of time in these experiments was set to one hundredth of the wake up time, $H_w$. To first illustrate the validity of the theoretical equations for describing the bubble in the limit of large number of agents, we superimpose in figure 6, the dynamics generated by the eqn. 18, onto the price history generated by the simulation for 40000 agents. The parameters used here are the same as those exhibited in figure 4. Again since the model does not incorporate any learning on the part of the agents, the dynamics generated is essentially periodic, and is presented this way only to highlight the agreement between the simulations and the theoretical equations.

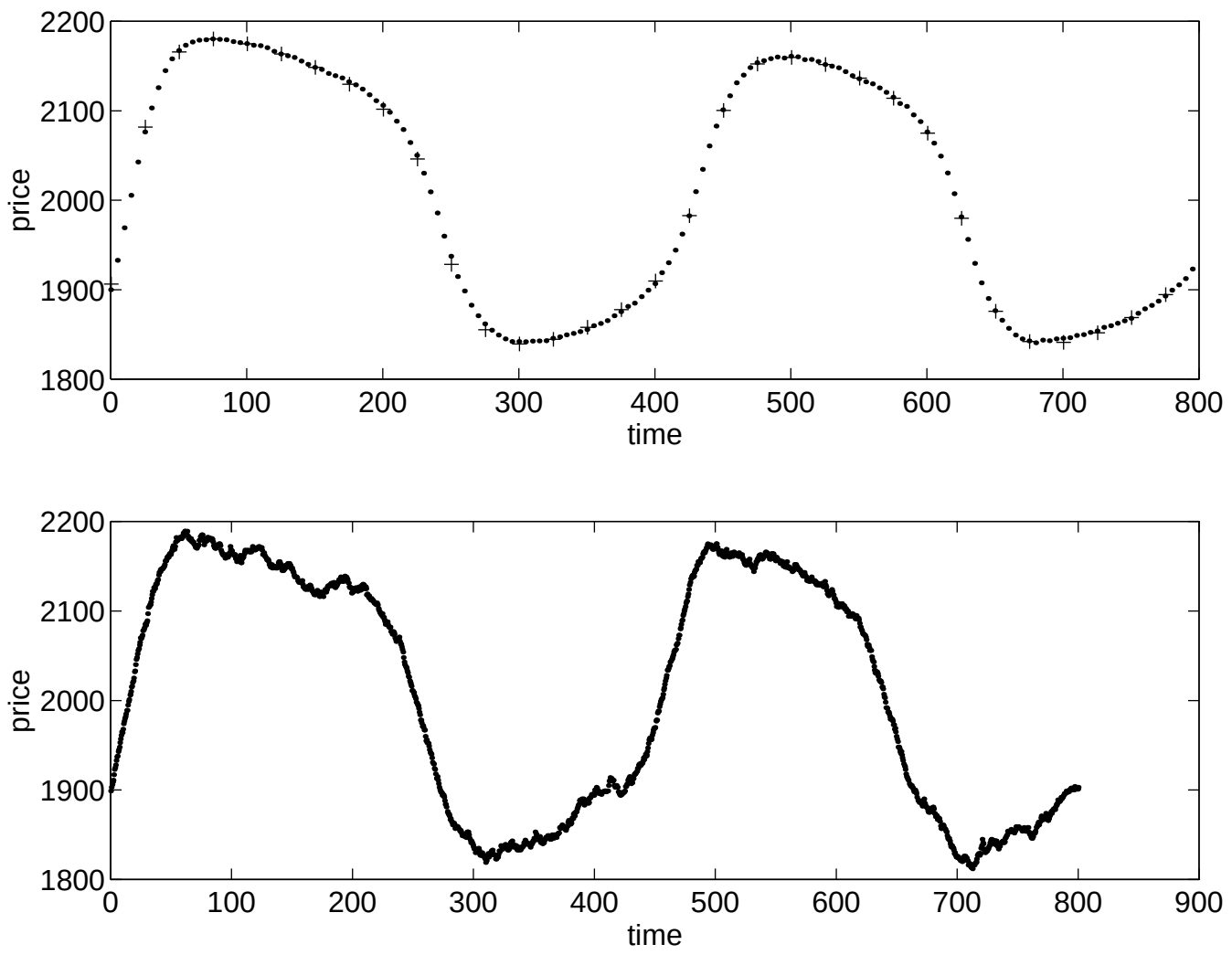

**Fig. 6.** Parameters same as for fig. 4. Top plot shows the results of the simulation for 40,000 agents (points) on top of which is superimposed results from eqns. (18) (crosses). Bottom plot shows the results of the simulation but this time with only 1,000 agents. Note how internally generated fluctuations are more pronounced when the dynamics are not in the trending regions.

Also shown in figure 6 is the price history generated for the same model, but this time with only 1000 agents. As can be seen, the agreement of the model the theory is only qualitative, since fluctuations that are suppressed in the large agent limit now play a role. It is important to notice that the dynamics is most sensitive to these fluctuations at the top and the bottom of the cycles. This increased sensitivity to fluctuations at the turning points is quite general and is also observed when these fluctuations are externally generated. This phenomenon is discussed further in the next section.

## Fluctuations

An interesting aspect of bubble dynamics has to do with the susceptibility of an inflated bubble to exogenous perturbations, such as sudden changes in systematic biases in agents expectations, or changes in the fundamental values. To examine the sensitivity of the bubble to the strength of such shocks, we assume that the fundamental value of all agents is lowered over some time period.

More specifically, in fig. 7, the fundamental values of all agents participating in the market was lowered by 50 price units for 50 time units after which fundamental values are returned to their original values. This fluctuation was successively applied in the simulations at times 425, 450, and finally at time 500. As can be seen, the former such fluctuation barely perturbs the bubble. The fluctuation at time 450 reverses the direction of price rises but the bubble soon continues to inflate. Finally, when the perturbation occurs at time 500, the bubble is weakened and an abrupt price drop occurs.

Depending on the timing of the shock the immediate consequence can range from negligible to drastic. In the former case the crash takes place in the same manner as without the shock, whereas in the latter the same perturbation leads to an induced crash which takes place earlier than that without the shock. This behavior can be understood in terms of figure 5 in the following fashion.

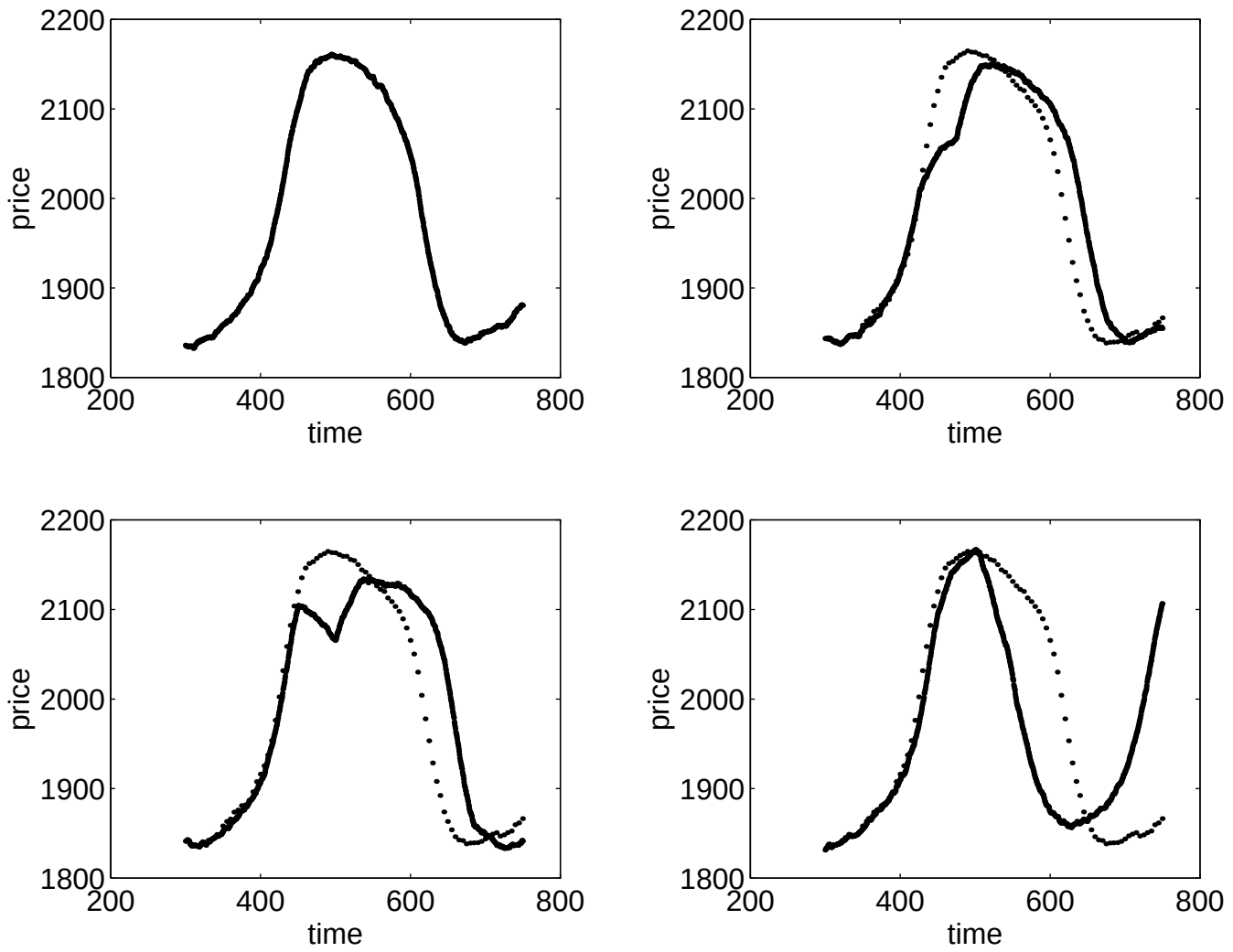

**Fig. 7.** Plots showing the varying susceptibility of a bubble to the timing of the same fluctuation. Upper left plot shows a simulation with 10,000 agents and with parameters as in fig. 4. Upper right plot shows the effect of a drop in agents' fundamental valuation by 50 price points lasting for 50 time units and starting at time 425. Lower left plot shows the effect of the same drop at time 450. Lower right plot shows the effect of the same drop at time 500. The susceptibility of the bubble to bursting thus increases as the bubble grows.

The price shock applied to the system in this manner has the effect of temporarily shifting the flow field of fig. 7 and then shifting it back after the shock. If during the shock, the shifted flow puts the system in a regime of falling prices, prices will fall. After, the shock, the flow field is shifted back to its original position, but the state of the system itself is has been changed. If the system is now in a regime of falling price, then the market price will fall, and otherwise it will continue to rise.

The plots in fig. 7 represent three cases. First, the case where the during the shock, the system state is still in a regime of rising prices. After the shock, the system is still in the region of rising prices in the flow field and the bubble continues to grow. Second, the case where during the shock, the system state is in a regime of falling prices. After the shock, however, the system state is again in the region of rising prices and the bubble begins to grow. Third and finally, during and after the shock the system state is in the region of falling prices and the bubble is burst.

## 5 Conclusion

In this paper, we have constructed a simplified model of bubble formation and bursting. A crucial ingredient of the theory is the explicit treatment of expectations on the part of speculative agents. We showed that when expectations are predominantly trend chasing, bubbles do form, with their subsequent crash. An analytic set of equation were derived by assuming a simplified version of market mechanisms. However, the model incorporates other important characteristics necessary to understand the essential features of bubbles, such as trend chasing and fundamental valuations.

Fundamental to the dynamics are the agent tendencies towards trend chasing on the one hand and the basing trades on fundamental factors. Price movements feed on themselves until a point is reached when some agents view the price increases as no longer sustainable, and trade instead more on the basis of fundamentals. This stops the bubble from expanding and is then followed by a decrease in prices. By considering the dynamics in the form of a flow field, we can qualitatively trace the “strength” of the bubble as its size increases.

In the context of this theory, it is easy to understand why the 1987 crash occurred even though there was no real news to justify the 22% devaluation on that day. Within the dynamical model presented in this paper, the crash did not represent such a huge devaluation in the fundamentals of the economy, but rather a situation in which trend chasing in the upward direction put the market in a critically overvalued situation. Trend chasing investment strategies on the part of institutional sellers then culminated in the crash itself.

There are several limitations to this model that will need to be overcome in order for it to become an accurate theory of bubble formation and market crashes. We have not incorporated the notion of variations in volume as determining current prices nor do we have any auction mechanisms to mediate the purchase or selling of assets. In this version, price increases and decreases caused by specific agents are assumed to scale as the inverse of the number of agents, and no allowance is made for diversity in the amount by which single agents influence market prices (they are all equally rich). As a result, short term fluctuations caused by these mechanisms do not appear in the dynamics. In spite of these constraints, this model points to departures from random walk behavior in the market that are bound to occur when large amounts of trend chasing are present.